\documentstyle[aps,prl,epsf,twocolumn]{revtex}
\begin{document}
\title{Approximate scaling relation for the anharmonic electron-phonon problem}
\author{J.\ K.\ Freericks$^1$, V. Zlati\'c$^{1,2}$, and M.\ Jarrell$^3$}
\address{$^1$Department of Physics, Georgetown University, Washington, DC 
20057--0995\\
$^2$Institute of Physics, Zagreb, Croatia\\
$^3$Department of Physics, University of Cincinnati, Cincinnati, OH 45221\\
}
\date{\today}
\maketitle
\begin{abstract}
An approximate scaling relation is found for the transition temperature to
a charge-density-wave instability in the anharmonic electron-phonon
problem, which maps a wide range of interaction strengths, anharmonicities,
and phonon frequencies onto a common functional form.  The relation 
employs the wave-function renormalization parameter and is valid even for 
systems that are not Fermi liquids.
\end{abstract}
\renewcommand{\thefootnote}{\copyright}
\footnotetext{ 1999 by the authors.  Reproduction of this article by any means
is permitted for non-commercial purposes.}
\renewcommand{\thefootnote}{\alpha{footnote}}

\pacs{Principle PACS number 74.20.-z; Secondary PACS numbers 63.20.Kr, 
63.20.Ry, 74.25.Dw}

\paragraph*{Introduction} The interaction of electrons with anharmonic 
lattice vibrations is a long-standing problem that is not yet fully understood.
What is surprising about this problem is that nearly all real materials
are anharmonic (as can be seen by the fact that they expand or contract 
upon heating), but quasiharmonic models (which replace the anharmonic
phonons by harmonic phonons with temperature-dependent phonon frequencies)
work remarkably well at describing properties of most materials 
\cite{quasiharm}.  
Superconductivity is described most accurately, where
the theory of electrons 
interacting with harmonic phonons, introduced by Migdal \cite{migdal} 
and Eliashberg \cite{eliashberg}, can routinely reproduce experimental
tunneling conductances
to better than one part in a thousand. The explanation for this result
is actually quite simple---the thermal effects that arise due to a nonuniform
spacing of the anharmonic energy levels, are unimportant when
the temperature is much less than the effective phonon frequency (defined by
the difference in energy between the ground and the first-excited state of
the anharmonic phonon) \cite{sofo_mahan}.  Furthermore, quantum Monte Carlo 
(QMC) studies \cite{qmc_anharm}, have shown that anharmonicity does not
appear to produce any exotic behavior, such as enhancements of transition
temperatures, or novel superconducting behavior \cite{hirsch_anharm}.
Instead, the results indicate that the main effect of anharmonicity is to 
generically
break particle-hole symmetry.  The discovery we present here is that
the anharmonic systems can be mapped onto harmonic ones, with
results from widely different parameter regimes collapsing onto the same
scaling curve.  We believe that this result sheds light onto the question
of why harmonic models work so well for describing properties of real
materials.

Our strategy is to solve anharmonic models
in the limit of large spatial dimension
\cite{metzner_vollhardt} where the lattice many-body problem can be mapped
onto a self-consistently embedded impurity problem that is solved via
a QMC simulation \cite{hirsch_fye} for quantum phonons, or via an iterative
transcendental equation for classical phonons \cite{brandt_mielsch,millis}.

\paragraph*{Model}
The simplest electron-phonon model that includes anharmonic effects
is the anharmonic Holstein model \cite{holstein} in which the conduction 
electrons interact with local phonon modes:
\begin{eqnarray}
H&=&-  \sum_{i,j,\sigma} t_{ij} c_{i\sigma}^{\dag }c_{j\sigma}
+ \sum_i (g \bar x_i - \bar\mu )(n_{i\uparrow}+n_{i\downarrow}) \cr
&+&\frac{1}{2M}\sum_i p_{i}^2+\frac{1}{2}\bar\kappa\sum_i\bar x_i^2+\alpha_{an}
\sum_i \bar x_i^4 \,.
\label{eq: ham}
\end{eqnarray}
Here, $c_{i\sigma}^{\dag}$ ($c_{i\sigma}$) creates (destroys) an electron at
site $i$ with spin $\sigma$, $n_{i\sigma}=c_{i\sigma}^{\dag}c_{i\sigma}$ is
the electron number operator, $\bar\mu$ is the chemical potential, and 
$\bar x_i$ ($p_i$) is the phonon coordinate
(momentum) at site $i$.  The hopping of electrons is restricted to
nearest-neighbor lattice sites $i$ and $j$ on a $d$-dimensional hypercubic
lattice.  $t_{ij}$ is isotropic with magnitude $t=:t^*/2\sqrt{d}$
where $t^*=1$ to define the energy scale.   The bare density of states then
becomes a Gaussian $\exp(-\epsilon^2)/\sqrt{\pi}$.  The local phonon has a 
mass $M$ and a spring constant $\bar\kappa$ associated with it.  The 
anharmonic contribution to the phonon potential energy is chosen 
to be a quartic in the phonon coordinate with a strength $\alpha_{an}$.  
The deformation potential (electron-phonon interaction strength) is
parameterized by an energy per unit length $g$.

A shift of the phonon coordinate is useful for calculations, and for 
illustrating the particle-hole symmetry of the model.  We shift $\bar x_i
=:x_i+x^{\prime}$,
with $g+\bar\kappa x^{\prime}+3\alpha_{an}x^{\prime 3}=0$, to transform the 
Hamiltonian into
\begin{eqnarray}
H&=&-  \sum_{i,j,\sigma} t_{ij} c_{i\sigma}^{\dag }c_{j\sigma}
+ \sum_i (g x_i - \mu )(n_{i\uparrow}+n_{i\downarrow}-1) \cr
&+&\frac{1}{2M}\sum_i p_{i}^2+\frac{1}{2}\kappa\sum_ix_i^2+\beta_{an}\sum_i
x_i^3+\alpha_{an} \sum_i x_i^4 ,
\label{eq: ham2}
\end{eqnarray}
with $\kappa:=\bar\kappa+12\alpha_{an}x^{\prime 2}$, $\beta_{an}:=4\alpha_{an}
x^{\prime}$,
and $\mu:=\bar\mu-gx^{\prime}$.  It is the presence of the cubic term when 
$\alpha_{an}\ne 0$ that removes particle-hole symmetry from the problem when
$\mu=0$, since the particle-hole transformation is $x_i\rightarrow-x_i$
and $n_{i\sigma}\rightarrow 1-n_{i\sigma}$.  
The system does continue to possess nesting at half-filling, though, which 
implies that it will have a nonzero transition temperature to a 
charge-density-wave (CDW) at half-filling for any nonzero interaction strength
$g$.

We examine two different cases in this contribution: quantum phonons with 
a large enough phonon frequency that vertex corrections are important
($\Omega=\sqrt{\kappa/M}=0.5$, corresponding to approximately one eighth
the effective bandwidth) and classical phonons with zero phonon frequency
($M\rightarrow\infty$ and $\Omega=0$).  The former problem is solved using
QMC techniques that have been described elsewhere \cite{qmc_anharm,hirsch_fye}.
Transition
temperatures are determined by calculating the relevant susceptibility (in this
case to a chessboard-phase CDW), and determining the temperature where it
diverges (when Trotter error is important, we extrapolate $T_c$ using 
$\Delta\tau=0.2$ and $0.4$, otherwise, we use the larger $\Delta\tau$).  
In the latter case, the problem is solved within the ordered phase,
employing a generalization of the Brandt-Mielsch formalism to the static
Holstein model \cite{brandt_mielsch,millis}, and determining $T_c$ as the 
highest temperature that sustains long-range order.

\paragraph*{Results} We begin in Fig.~1(a)
by showing the transition temperature to the
commensurate CDW at half filling for the harmonic case, and three different
phonon frequencies, $\Omega=0$, $\Omega=0.5$, and $\Omega=\infty$ which 
is identical to the attractive Hubbard model \cite{qmc_anharm,jarrell_hubbard}.
The symbols are the exact, or QMC results (the dotted lines are guides to the 
eye), while the other lines are second-order
weak-coupling conserving approximations \cite{holst_pert}
(dashed) and second-order strong-coupling calculations \cite{holst_strong}
(solid).  Notice how the maximal $T_c$ is essentially independent of
phonon frequency, and that the phase diagrams are not too sensitive to
phonon frequency when $\Omega$ is smaller than the bandwidth.  However, note
that the classical phonon case ($\Omega=0$), and the quantum phonon case are 
quite different,
since $T$ is always above $\Omega$ for the classical phonons, but $T<\Omega$
for the quantum phonon cases shown here. The similarity in $T_c$ is surprising, 
because the CDW vertex is strongly temperature dependent for the quantum
phonons, with it's magnitude changing between $-2g^2/\kappa$ and $-g^2/\kappa$
as the temperature and frequency are varied.

The anharmonicity is turned on in Fig.~1(b) for similar effective
electron-electron coupling strengths and $\Omega=0$ and $\Omega=0.5$.  The
infinite-phonon frequency limit is unaffected by the anharmonicity, since the
phonons respond instantaneously to the electrons, and remain at the origin.
We plot the strength of the anharmonicity as $\alpha_{an}\bar 
x^{*4}$, where $\bar x^*$
is the equilibrium coordinate of the atomic system with one electron per site,
found from Eq.~(\ref{eq: ham}) with $t_{ij}=0$.  In the case of weak
anharmonicity, this equilibrium coordinate lies near $-g/\kappa$ and the
anharmonic energy grows linearly with $\alpha_{an}$, but as the anharmonicity
becomes large, then $\bar 
x^*$ lies near $-(g/\alpha_{an})^{1/3}$, and the anharmonic
energy decreases as $\alpha_{an}^{-1/3}$.  This is why the curves in Fig.~(1(b)
are multivalued for some values of the interaction strength.  Notice that 
the results for different phonon frequencies are similar in qualitative 
behavior, but that the quantitative results can differ by large amounts
as the anharmonicity increases.

We find that the $T_c$ satisfies an approximate scaling relation when we plot 
it instead
as a function of $1/Z(0)$, the wave-function renormalization parameter,
in Fig.~2.  Nearly two-hundred data points collapse onto the same scaling 
curve.  But the scaling is only approximate, because the infinite-phonon 
frequency limit, lies well off the scaling curve for low-to-moderate phonon 
frequencies, so this result must break down as $\Omega$ increases (note, 
however, that the value $\Omega/t^*=0.5$ is larger than the phonon frequency
in nearly all real materials).  The other 
lines in Fig.~2 correspond to different weak-coupling approximations plotted 
now as a function of $Z(0)$ rather than of $U$.  The wave function 
renormalization parameter $Z(0)$ is extracted from the calculations by a 
linear extrapolation along the Matsubara frequency axis.  We compute
\begin{equation}
Z(0)=1-\frac{3}{2}\frac{{\rm Im}
\Sigma(i\pi T)}{\pi T}+\frac{1}{2}\frac{{\rm Im}\Sigma( 3i\pi T)}{3\pi T},
\label{eq: extrap}
\end{equation}
and evaluate it at a temperature $T$ just above $T_c$.  This imaginary-axis
extrapolation procedure is 
robust in producing a scaling result even if the system is not a Fermi-liquid, 
where $1/Z(0)$ would measure the quasiparticle weight.  For example, in the 
classical-phonon case, the lowest-order contribution to the self energy is 
$\Sigma(z)=-TUG(z)$.
The curvature of ${\rm Im}\Sigma(\omega)$ at the chemical potential is 
positive.  Hence the system is never a Fermi liquid, except at $T=0$, where 
it becomes noninteracting. 
Nevertheless, the extrapolation procedure given in 
Eq.~(\ref{eq: extrap}) still falls on the scaling curve, and for the 
weakly-coupled classical-phonon case, it produces $Z(0)=1+4|U|\rho(0)/3$, 
rather than the expected result of $Z(0)=1+|U|\rho(0)$ from Migdal-Eliashberg 
theory. We find that a phenomenological functional form that fits the 
data relatively well is $T_c=0.182 Z(0)^{-0.12}\exp[-0.75/\{Z(0)-1\}]$  
(the solid line in Fig.~2).

As a consequence of this scaling, if we use the quasiharmonic approximation to 
describe an anharmonic system with the parameters adjusted to obtain 
correspondence in $Z(0)$, then we also accurately reproduce two-particle
properties like $T_c$.  Since this result is robust against introducing
anharmonicity, we believe this is the reason why the quasiharmonic 
approximation works so well in real materials.

We plot the interacting density of states and the self energy in Figs.~3-5
for weak-coupling harmonic and anharmonic cases, and for a strong-coupling
harmonic case.  We range from a high temperature down to just above $T_c$.
Notice how the weak-coupling cases are not Fermi liquids, and have quite
different DOS and self energies, but both map onto the same $Z(0)$ and $T_c$
[our definition of $Z(0)$ from the imaginary axis is not equal to
the derivative of the self energy on the real axis, which has the opposite 
sign for a Fermi liquid here].
Furthermore, the paramagnetic phase of this model becomes noninteracting as
$T\rightarrow 0$ since $\Sigma$ vanishes in the weak-coupling limit, which
also explains why the anharmonic DOS becomes more symmetric as $T\rightarrow 0$.
Scaling even holds (in a more approximate way) when the system is a bipolaronic
insulator as shown in Fig.~5---the self-energy has a large imaginary part near 
the chemical potential, which grows as $T$ is lowered, and a pseudogap develops
in the interacting density of states.  This is then followed by a gap developing
at lower $T$ and the imaginary part of the
self-energy vanishing within the gap region except for a narrow
spike at the frequency where the real part of the self-energy
changes sign.  But the $Z(0)$
parameter can still be defined on the imaginary axis!

\paragraph*{Conclusion}

We have discovered an approximate scaling relation for the electron-phonon
problem that holds over a wide range of phonon frequencies, coupling strengths,
and anharmonicities.  It relates the wavefunction renormalization parameter,
extracted from the imaginary axis, to the transition temperature.  Our
results show that anharmonic models that are tuned to the same $Z(0)$ will 
show similar $T_c$s as harmonic models with the same $Z(0)$.  We feel this
underlying scaling behavior helps explain the success of the quasiharmonic
approximation employed in describing nearly all real materials.  

We would like to acknowledge useful conversations with 
J.\ Hirsch, 
G.\ Mahan,
P. Miller, 
B. Sch\"uttler,
and J.\ Serene.
J.~K.~F. acknowledges support of ONR grant
YIP-N000149610828 and J.~K.~F. and V.~Z. 
acknowledge support of NSF grants DMR-9627778 and INT-9722782. 
M.~J. acknowledges support of NSF grants DMR-9704021 and DMR-9357199.
Supercomputer time was provided by grant DMR950010P from 
the Pittsburgh Supercomputing Center, sponsored by the NSF.


\begin{figure}[ht]
\caption{Transition temperatures for the chessboard phase charge-density-wave
order at half filling.  (a) $T_c$ for the harmonic case.  The horizontal
axis is the effective electron-electron coupling $|U|=g^2/\kappa$.  Three
frequencies are included $\Omega=0$, $\Omega=0.5$, and $\Omega=\infty$.  The
dashed line is a conserving second-order perturbative approximation (including
vertex corrections) and the solid line is a second-order strong-coupling
approximation. (b) $T_c$ for the anharmonic case.  The horizontal axis is
a measure of the anharmonic contribution to the energy (as described in the 
text). Open symbols are QMC simulations at $\Omega=0.5$, while solid symbols
are exact solutions at $\Omega=0$.  
}
\end{figure}

\begin{figure}[ht]
\caption{Scaling curve for the anharmonic phonon problem.  The horizontal
axis is the imaginary-axis-extrapolated wave-function renormalization
parameter, extrapolated at $T_c$, while the vertical axis is the CDW $T_c$
at half filling.  Included in the curve are
weak-coupling results for the classical phonons with (dotted) 
vertex corrections, a conserving second-order approximation
for the $\Omega=\infty$ case (dashed), and the 
approximate form described in the text (solid).}
\end{figure}

\begin{figure}[ht]
\caption{Interacting density of states for a generic weak-coupling harmonic case
($\Omega=0$,
$g=0.625$ and $\alpha_{an}=0.0$) for six temperatures running from top to 
bottom at $\omega=0$ $T=0.04$, 0.0625, 0.125, 0.25, 0.5, 1.0, and 2.0.  
The real and imaginary parts of the 
self energy are plotted in the insets.  Notice how the slope of the real
part of $\Sigma$ has the wrong sign, and how the imaginary part has a local
minimum at the chemical potential, signifying that
this system is not a Fermi liquid.}
\end{figure}

\begin{figure}[ht]
\caption{Interacting density of states for a generic weak-coupling anharmonic
case ($\Omega=0$,
$g=2.5$ and $\alpha_{an}=4.5$) for the same six temperatures 
shown in Figure 3.  Notice how the density of states is
now asymmetric at high $T$ but becomes more symmetric as $T$ decreases. }
\end{figure}

\begin{figure}[ht]
\caption{Interacting density of states for a generic strong-coupling anharmonic
case ($\Omega=0$,
$g=2.5$ and $\alpha_{an}=0.004$) for six temperatures running from top to
bottom (at $\omega=0$) $T=0.075$, 0.125, 0.25, 0.5, 1.0 and
2.0. Notice how the density of states 
develops a pseudogap at the chemical potential, 
which then becomes a true gap as $T$ is lowered.}
\end{figure}

\begin{figure}[ht]
\epsfxsize=2.5in
\epsffile{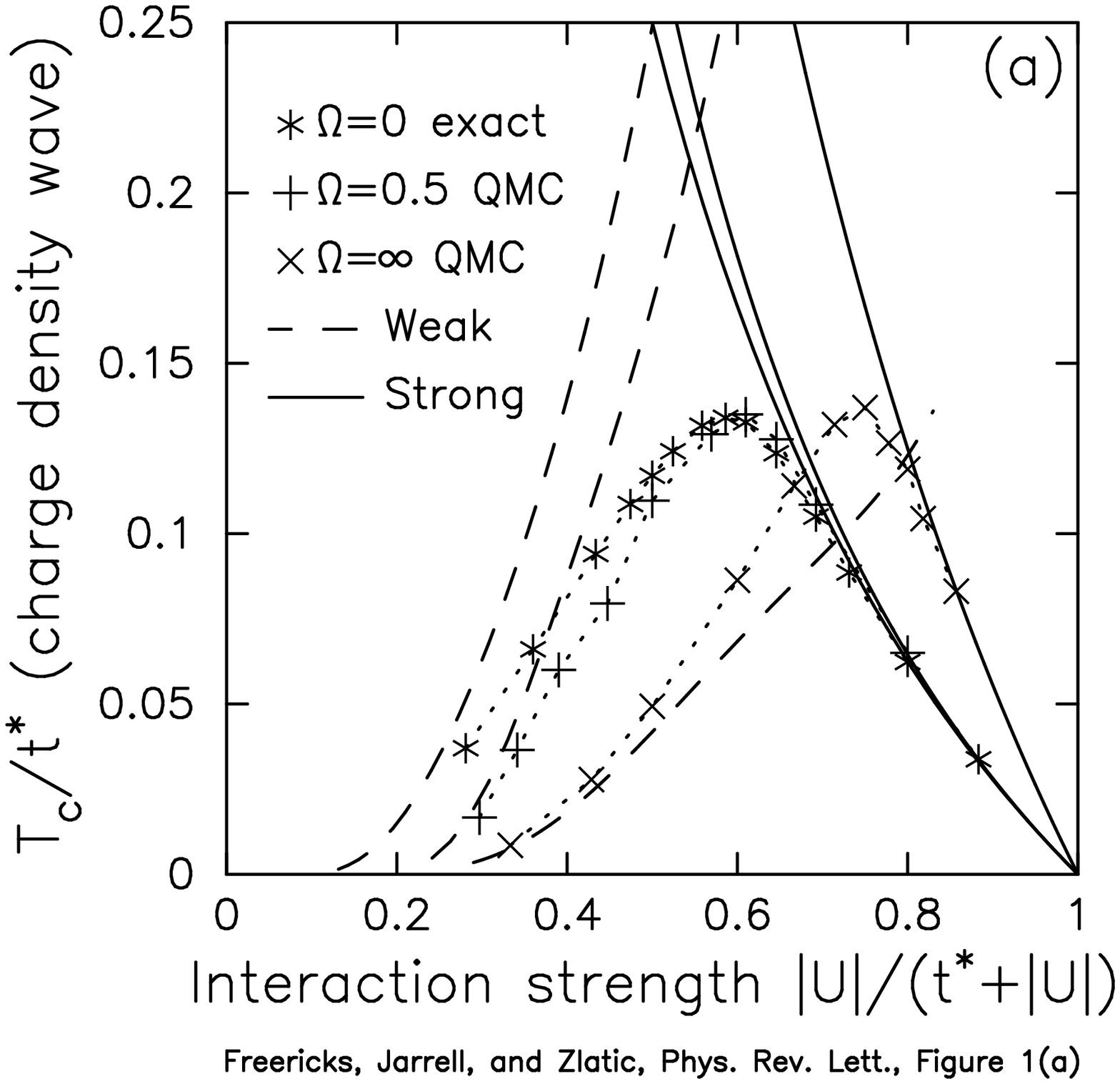}
\end{figure}

\begin{figure} [ht]
\epsfxsize=2.5in
\epsffile{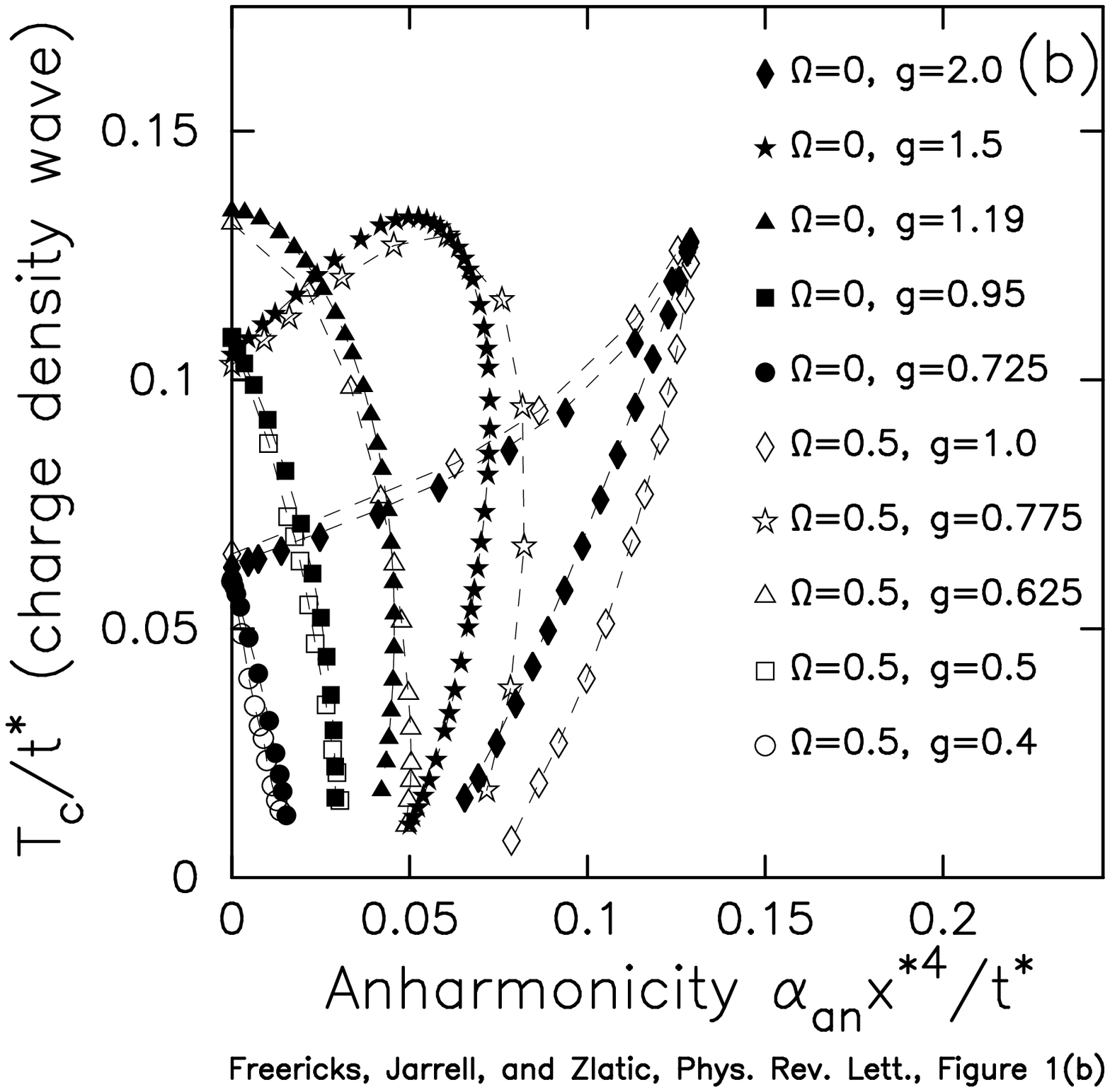}
\end{figure}

\begin{figure}[ht]
\epsfxsize=2.5in
\epsffile{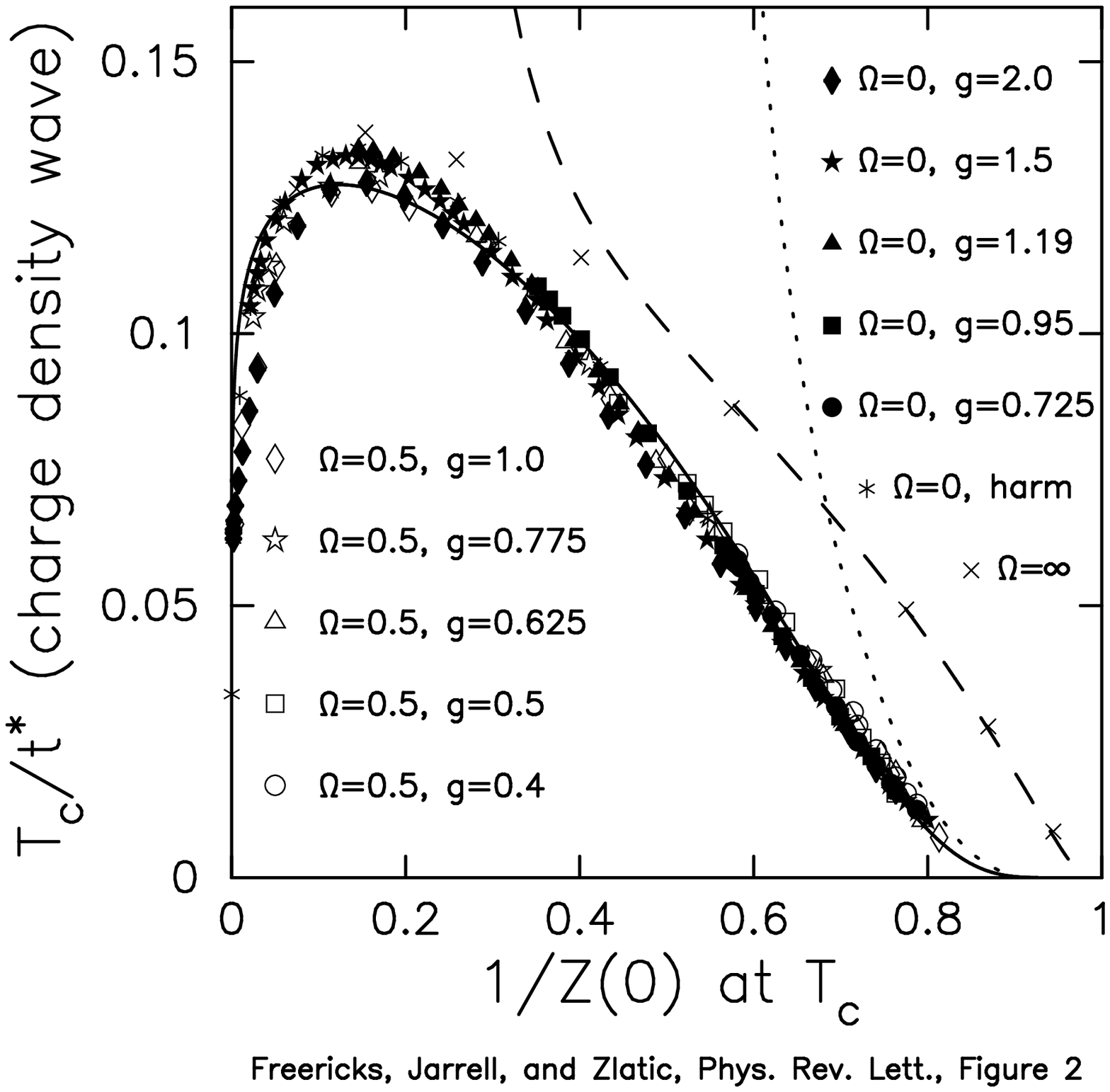}
\end{figure}

\begin{figure}[ht]
\epsfxsize=2.5in
\epsffile{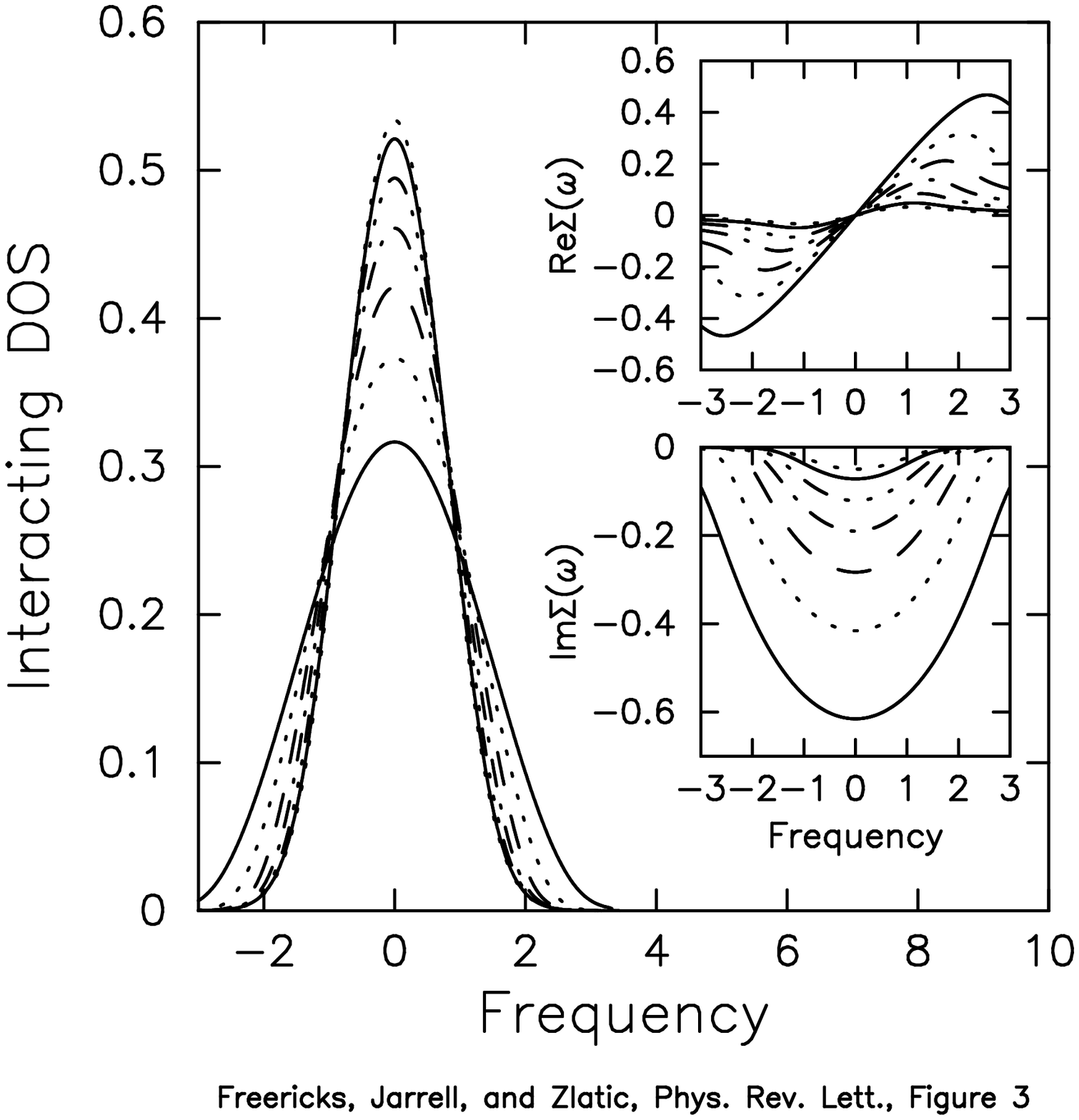}
\end{figure}

\begin{figure}[ht]
\epsfxsize=2.5in
\epsffile{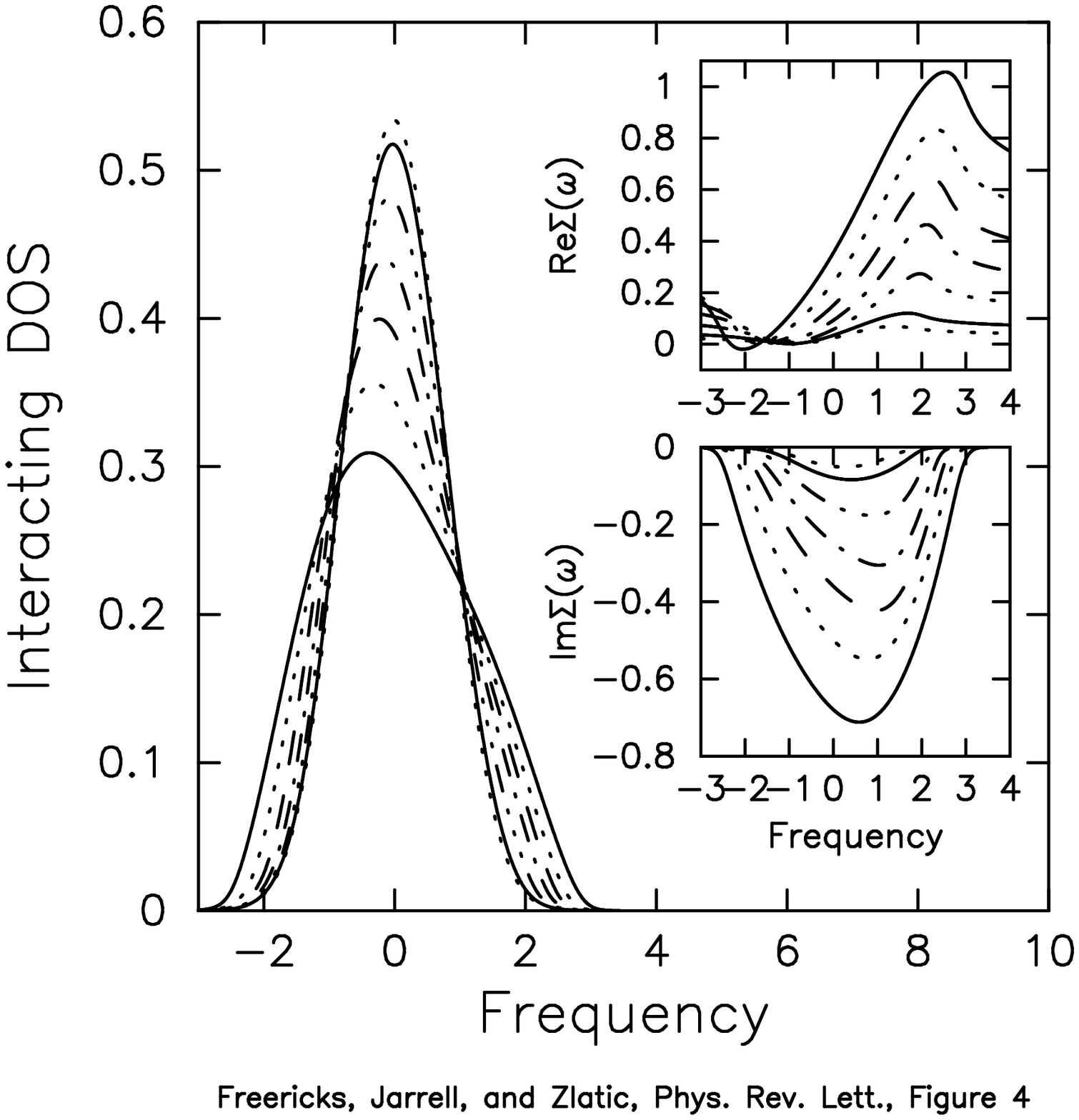}
\end{figure}

\begin{figure}[ht]
\epsfxsize=2.5in
\epsffile{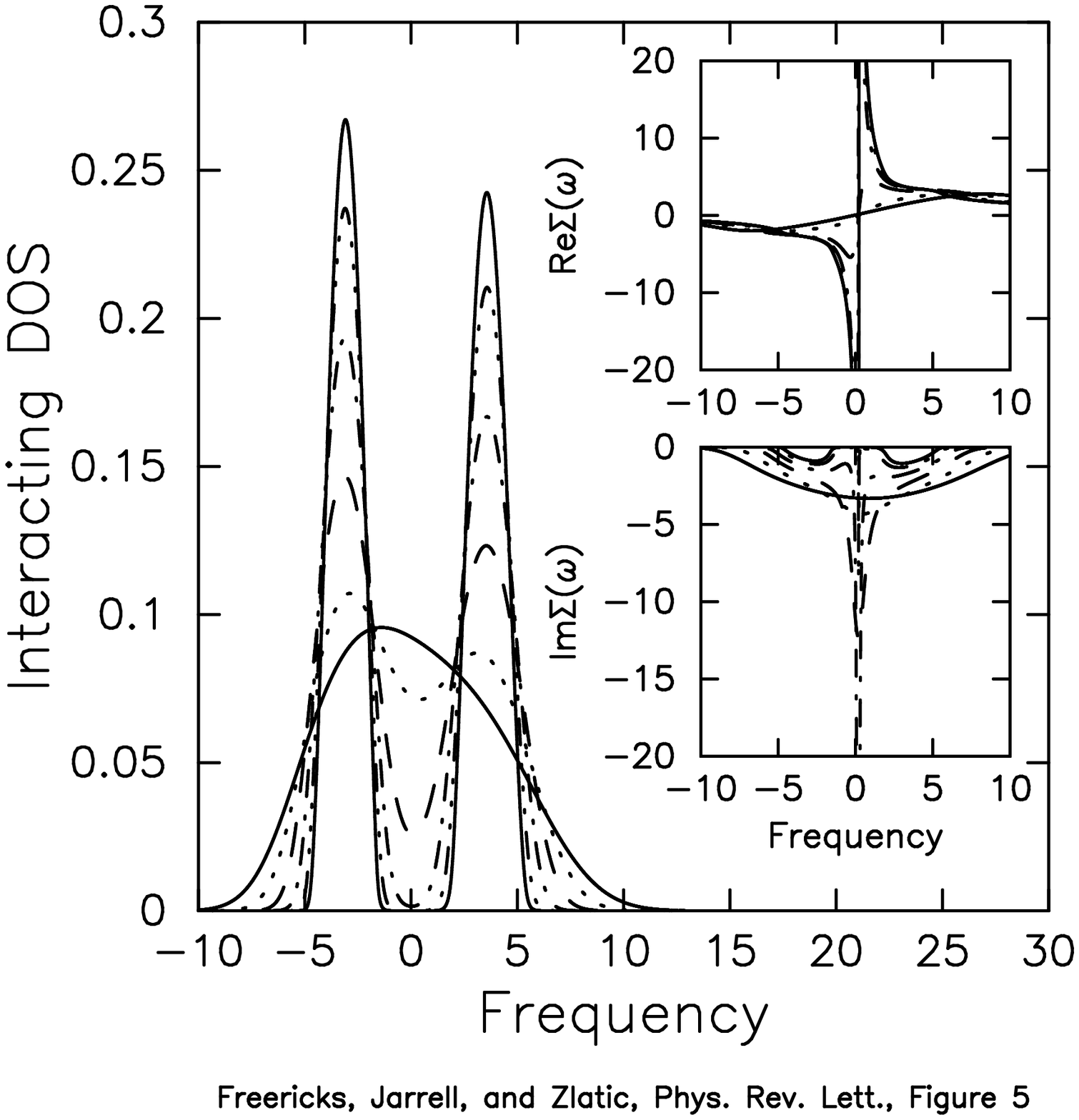}
\end{figure}

\end{document}